\documentclass[aps,prb,twocolumn,amsmath,amssymb,floatfix,superscriptaddress]{revtex4}

\usepackage{graphicx}
\usepackage{epsfig}
\usepackage{dcolumn}

\newcommand{\I}{\mathrm{i}}

\newcommand{\C}{\mathrm{C}}
\newcommand{\leads}{\mathrm{leads}}

\begin{document}

\title{ Spin qubits with electrically gated polyoxometalate molecules
}

\author{J\"org Lehmann}
\affiliation{Department
of Physics und Astronomy,
University of Basel,
Klingelbergstrasse~82,
CH-4056~Basel, Switzerland}
\author{Alejandro Gaita-Ari{\~n}o}
\affiliation{Department
of Physics und Astronomy,
University of Basel,
Klingelbergstrasse~82,
CH-4056~Basel, Switzerland}
\affiliation{
Instituto de Ciencia Molecular,
Universitat de Valencia,
Pol\'{i}gono de La Coma, s/n,
E-46980~Paterna, Spain}
\author{Eugenio Coronado}
\affiliation{
Instituto de Ciencia Molecular,
Universitat de Valencia,
Pol\'{i}gono de La Coma, s/n,
E-46980~Paterna, Spain}
\author{Daniel Loss}
\affiliation{Department
of Physics und Astronomy,
University of Basel,
Klingelbergstrasse~82,
CH-4056~Basel, Switzerland}

\date{\today}

\begin{abstract} Spin qubits offer one of the most promising routes to
the implementation of quantum computers.  Very recent results in
semiconductor quantum dots show that electrically-controlled gating
schemes are particularly well-suited for the realization of a universal
set of quantum logical gates.  Scalability to a larger number of qubits,
however, remains an issue for such semiconductor quantum dots.  In
contrast, a chemical bottom-up approach allows one to produce identical
units in which localized spins represent the qubits.  Molecular
magnetism has produced a wide range of systems with tailored properties,
but molecules permitting electrical gating have been lacking.  Here we
propose to use the polyoxometalate [PMo$_{12}$O$_{40}$(VO)$_2$]$^{q-}$,
where two localized spins-$\mathbf{1/2}$ can be coupled through the
electrons of the central core.  Via electrical manipulation of the
molecular redox potential, the charge of the core can be changed. With
this setup, two-qubit gates and qubit readout can be implemented.
\end{abstract}

\pacs{}

\maketitle

\section{Introduction}

Quantum dots have often been termed artificial atoms and
molecules.\cite{vanderWiel2003a}  Indeed, their tunability via
electrical gates has permitted one to reach into a regime where only a
single electron sits in each dot.  A double dot then becomes the
analogue of a hydrogen molecule.  The Heisenberg exchange coupling in
such a system can be manipulated by appropriately chosen electrical
pulse sequences, up to a point which has not yet been achieved for its
real counterpart.\cite{Burkard1999a} Recently, it has been
demonstrated that this technique allows the realization of the
fundamental one- and two-qubit quantum
gates~\cite{Barenco1995a,Loss1998a,Nielsen2000a} in such a
system.\cite{Petta2005a,Koppens2006a}

Naturally, the question arises whether the successful schemes and
techniques of an electrical control of spins in quantum dots can be
transfered back to electron spins in single molecules. In view of the
recent experimental progress in the field of molecular electronics,
which demonstrates that magnetic transitions in atomic
chains~\cite{Hirjibehedin2006a} and single
molecules~\cite{Heersche2006a, Jo2006a} can be resolved in transport
measurements, such a goal should in principle be achievable. In
particular, the field of molecular
magnetism~\cite{Kahn1993a,Gatteschi2006a} has provided a plethora of
systems of almost arbitrary magnetic
functionality.\cite{Wernsdorfer1999a,Coronado2000a,Real2003a,Stepanow2004a,
  ChemRev} Very recently, phase-coherence times of up to $3\,\mu s$
have been reported for such molecular nanomagnets after
deuteration.\cite{Ardavan2007a} Yet, a suitable system for an
electrical control of molecular qubits has still been missing.  It is
this electrical control which is crucial for scalability and which
distinguishes this proposal from earlier ones also based on molecular
magnets.~\cite{Leuenberger2001a,Troiani2005a}

\begin{figure}[ht]
  \centering
  (a) \raisebox{-4.5cm}{\includegraphics[width=0.4\linewidth]{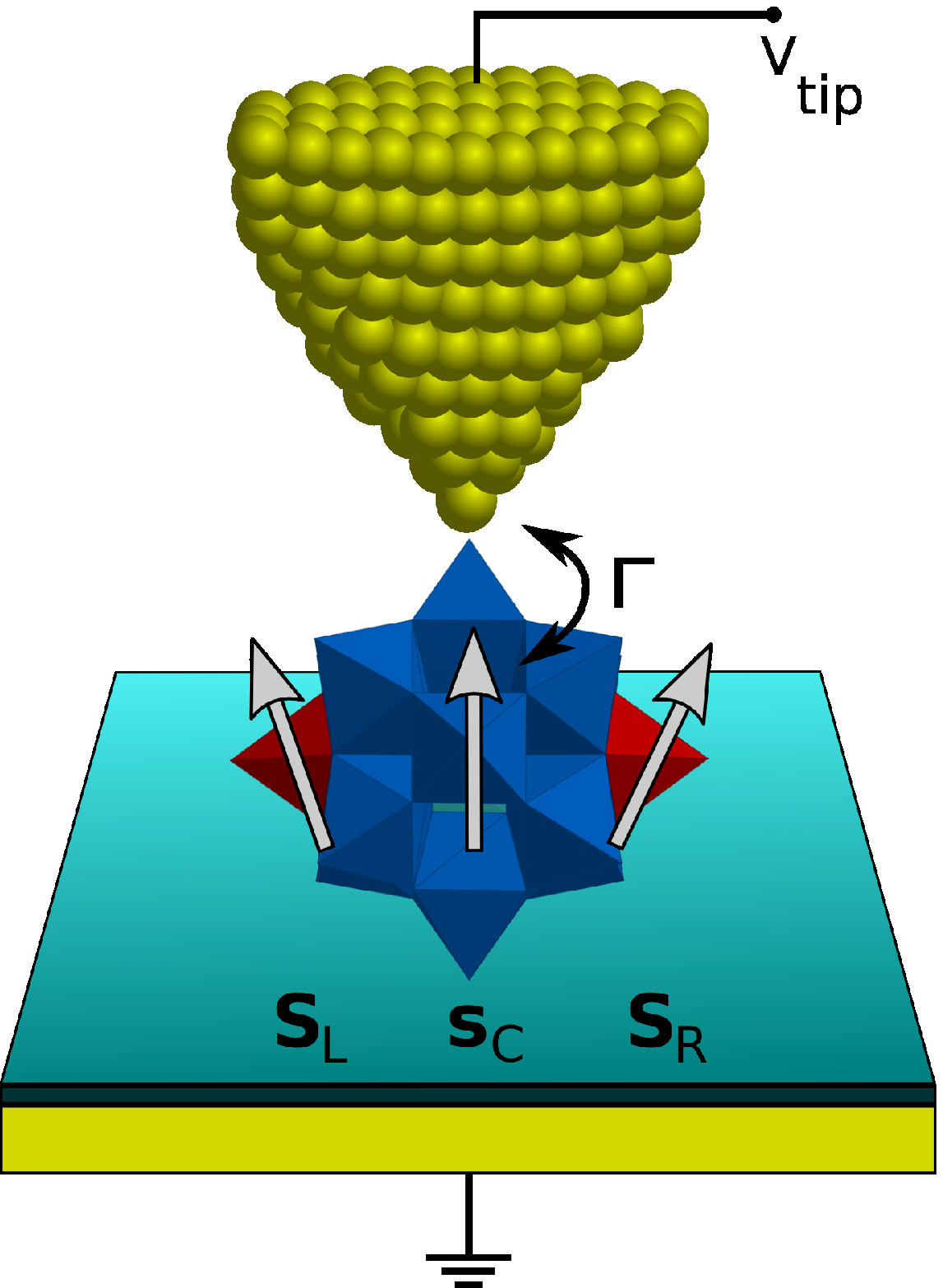}}
  (b) \raisebox{-3.7cm}{\includegraphics[width=0.45\linewidth]{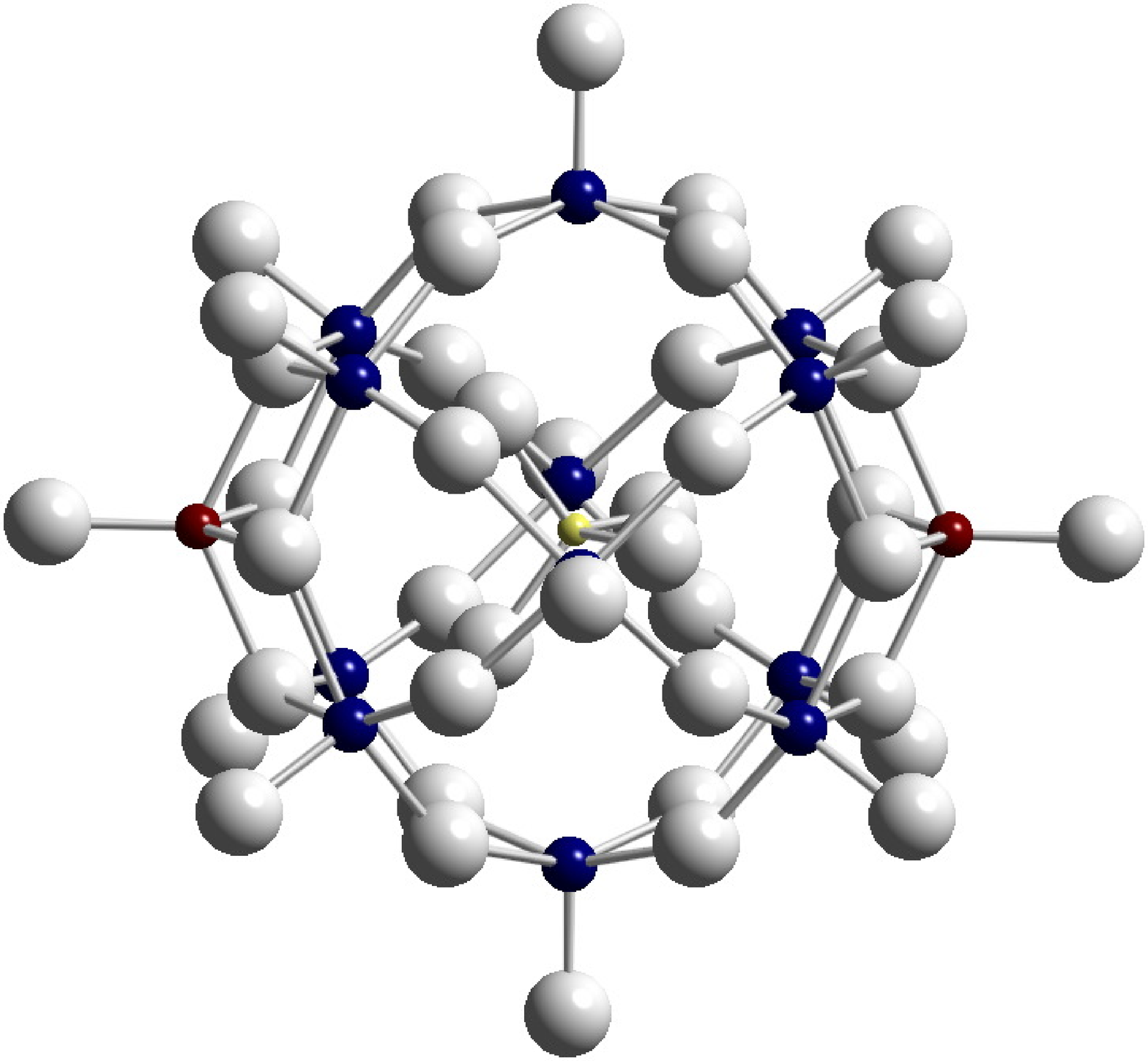}}
  \caption{(a) Schematic drawing of the polyoxometalate
    [PMo$_{12}$O$_{40}$(VO)$_2$]$^{q-}$ separated by a tunnelling
    barrier from a metallic substrate and contacted via a tunnel
    coupling~$\Gamma$ to a tip at a potential $V_\mathrm{tip}$.
    Indicated are the two localized spins $\mathbf{S}_\mathrm{L}$ and
    $\mathbf{S}_\mathrm{R}$ of the V atoms in the center of the red
    square pyramids. Depending on the redox state of the molecule, the
    delocalized valence electrons of the Mo atoms in the center of the
    blue octahedra form a spin $\mathbf{S}_\C$ or pair to a singlet.
    The O anions are located at the vertices of the polyhedra.  (b)
    Ball-and-stick model of the polyoxometalate: O (grey), Mo (blue),
    V (red), and P (yellow).}
  \label{fig:model}
\end{figure}

Here we propose an experimental setup which permits the electrical
switching of the exchange interactions between two electron spins. The
system we choose for illustrating our proposal is
[PMo$_{12}$O$_{40}$(VO)$_2$]$^{q-}$,~\cite{Chen1996} a
polyoxometalate~\cite{ChemRev} which consists of a central
mixed-valence core based on the [PMo$_{12}$O$_{40}$]
Keggin~\cite{Keggin1933a} unit, able to act as an electron reservoir
accommodating a variable number of delocalized electrons hopping over
the Mo centers, capped by two vanadyl groups containing two localized
spins (cf.\ Fig.~\ref{fig:model})
The spins on these two (VO)$^{2+}$ units are weakly magnetically
coupled via the delocalized electrons of the central core.  We shall
show how this magnetic coupling can be switched in an all-electrical
way and how this can be used for the implementation of a fundamental
two-qubit gate (providing entanglement), the so-called
square-root-of-swap $\sqrt{\text{\sc SWAP}}$.~\cite{Loss1998a}
Furthermore, we will detail how to use a variation of this procedure
for the readout of the final state of the two qubits.

\section{Low-energy model of the polyoxometalate}

For the description of the low-energy states of the polyoxometalate,
two cases have to be distinguished: For an \textit{even} number of
electrons on the mixed-valence Keggin core, their spins pair
antiferromagnetically to form a total spin~$0$ state. Then the system
can be modelled by the two spins $1/2$ on the vanadyl groups weakly
coupled via an indirect exchange mechanism mediated by the core
electrons. On the other hand, if the number of core electrons is
\textit{odd}, an unpaired spin $1/2$ on the core remains and one
obtains a set of three coupled spins $1/2$.  Restricting ourselves to
two charge states differing by one electron, we can choose the
electron number of the even state as reference and write the
Hamiltonian in the form
\begin{equation}
  \label{eq:H}
  \begin{split}
    H = {} & -  J(n_\C)\,
    \mathbf{S}_\mathrm{L}\cdot \mathbf{S}_\mathrm{R} - J_\C
    (\mathbf{S}_\mathrm{L} + \mathbf{S}_\mathrm{R}) \cdot
    \mathbf{s}_\C
    \\ & 
    + (\epsilon_0 - e V_\mathrm{g})\, n_\C
    + U n_\C(n_\C-1)/2
    \,.
  \end{split}
\end{equation}
Here, the first term describes the indirect-exchange coupling between
the left and right spins $\mathbf{S}_\mathrm{L}$ and
$\mathbf{S}_\mathrm{R}$ with a coupling constant~$J(n_\C)$ which
depends on the number of electrons $n_\C$---measured with respect to
the reference number---on the central core of the molecule. In the
case of an electron being localized on the core, its
spin~$\mathbf{s}_\C = (1/2)\sum_{\sigma\sigma'} d^\dagger_{\C\sigma}
\,\boldsymbol{\tau}_{\sigma\sigma'}\, d_{\C\sigma'}$ couples to the
left and right spins with coupling constant~$J_\C$. Here, the
operators $d_{\C,\sigma}$ ($d^\dagger_{\C,\sigma}$) destroy (create)
and electron on the central core and $\boldsymbol\tau$ is the vector
of the three Pauli matrices.  The two last terms contain the orbital
energy~$\epsilon_0$ of the electron on the central core, which can be
shifted due to a gate potential~$V_\mathrm{g}$, and the molecule's
charging energy~$U$, respectively. The latter is assumed to be the
largest energy scale of the problem. We consider a situation where the
central core of the molecule is tunnel-coupled to one or more metallic
leads~$\ell$ permitting electrons to flow on and off the molecule with
tunnelling rates $\Gamma_\ell$. In the STM setup depicted in
Fig.~\ref{fig:model}(a), two leads, the tip and the metallic surface,
are present. We assume that an insulating layer---which still allows
electrons to tunnel through---between the surface
and the molecule leads to a molecule-surface coupling which is much
smaller than the typical energy scale~$J_\C$ of the molecular
Hamiltonian~\eqref{eq:H}. An applied bias
voltage~$V_\mathrm{b}=V_\mathrm{tip}$ then leads to a shift of the
molecular levels which to a very good approximation is described by
the linear relation $V_\mathrm{g}= \eta V_\mathrm{tip}$ with
$0<\eta<1$.~\cite{Datta1997a} Note that in a more sophisticated
three-terminal setup, the parameters $V_\mathrm{g}$ and $V_\mathrm{b}$
can be controlled independently.~\cite{Park1999a,Kubatkin2003a}

\section{Ideal quantum-gate operation}

\begin{figure}[ht]  
  \includegraphics[width=0.9\linewidth]{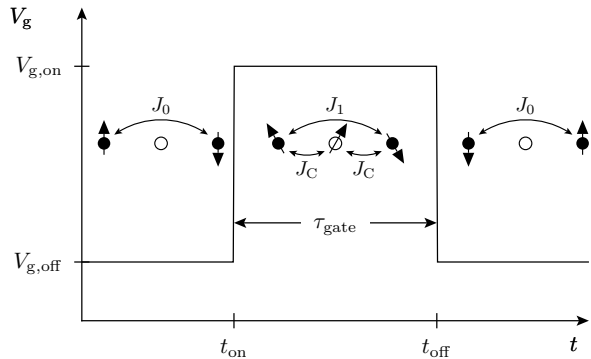}
  \caption{Quantum-gate sequence for the
    $\sqrt{\mathrm{SWAP}}$ operation and exchange coupling constants
    during the corresponding gate phase. The gate is turned on for a
  time $\tau_\mathrm{gate}.$}
  \label{fig:gating_sequence}
\end{figure}

The $\sqrt{\mathrm{SWAP}}$ operation is defined by
$|\chi,\lambda\rangle \rightarrow \left( | \chi,\lambda\rangle +
  \mathrm{i}|\lambda,\chi\rangle\right)/(1+\mathrm{i})$, where
$\chi,\lambda = \uparrow, \downarrow$. For electrically confined
quantum dots, it can be physically implemented by an appropriately
chosen gate-voltage pulse which turns on the Heisenberg exchange coupling
between the qubits for a specific time (see below).\cite{Loss1998a}
In a molecular system, however, the exchange constants are fixed by
the chemical structure and cannot be directly changed. Thus, a more
sophisticated quantum-gate scheme has to be employed. We propose to
indirectly manipulate the coupling between the two qubits by changing
the occupation~$n_\C$ of the central core in the molecule described
above. This change in $n_\C$ can be induced by changing the gate
voltage $V_\mathrm{g}$, such that different charge sectors of the
molecule become stable (cf.\ Fig.~\ref{fig:gating_sequence}).
Chemically this amounts to a change in the redox potential of the
molecule. We will first discuss the basic idea of the quantum-gate scheme,
thereby assuming that the electron number~$n_\C$ can be changed in a
deterministic, externally controllable way. Later on, this assumption
will be relaxed and the full tunnelling dynamics due to the change in
$V_\mathrm{g}$ will be included.

Initially, one adjusts the gate voltage~$V_\mathrm{g}$ in such a way
that the stable configuration is given by an even number of electrons
on the central core ($n_\mathrm{C}=0$). We assume that the exchange
coupling $J_0=J(0)$ between the spins $\mathbf{S}_\mathrm{L}$ and
$\mathbf{S}_\mathrm{R}$ will then be very small and can be disregarded
for times much smaller than $\hbar/J_0$ (see also below).  The quantum-gate
operation begins by changing the gate voltage such that the
$n_\mathrm{C}=1$ charge sector becomes stable. Then, on a time-scale
of the inverse tunnelling rate, an electron enters the central core.
After this has happened, the dynamics of the three-qubit system is
governed by the Hamiltonian~\eqref{eq:H} in the $n_\C=1$ subspace. The
spin-dependent part of this Hamiltonian becomes
\begin{equation}
  \label{eq:H1_t}
  H_1 =   - (J_1-J_\C)\,
  \mathbf{S}_\mathrm{L} \cdot \mathbf{S}_\mathrm{R}
  -\frac{J_\C}{2} \,\mathbf{S}^2 
  \,,
\end{equation}
where
$\mathbf{S}=\mathbf{S}_\mathrm{L}+\mathbf{S}_\mathrm{R}+\mathbf{s}_\C$
is the total spin of the system and $J_1=J(1)$. Before considering the
time-evolution due to this Hamiltonian, we conclude our discussion of
the gate cycle: after a specific time~$\tau_\mathrm{gate}$, one
switches back to the initial gate voltage and the excess electron
tunnels off the central core again. After this has happened, the two
outer spins are, up to times much smaller than $\hbar/J_0$, decoupled
again.

The only non-trivial dynamics is induced by the
Hamiltonian~\eqref{eq:H1_t}. The first term contains an effective
exchange coupling between the two spins. Up to an irrelevant phase
factor, its corresponding time-evolution yields the
$\sqrt{\mathrm{SWAP}}$ gate if the gate time $\tau_\mathrm{gate}$
fulfils~\cite{Loss1998a}
\begin{equation}
  (J_1-J_\C) \, \frac{\tau_\mathrm{gate}}{\hbar} = \frac{\pi}{2} +
  2\pi \, n\,,
  \label{eq:tswap}
\end{equation}
with $n$ being an arbitrary integer. The second term in
equation~\eqref{eq:H1_t} depends on the total spin of the three-spin-$1/2$
system. While this quantity is unknown, we can eliminate its influence
on the gate behavior if we restrict ourselves to stroboscopic gate
times $\tau_\mathrm{gate}$ for which the contribution
$\exp\left[\mathrm{i} (J_\C/2)\mathbf{S}^2 \tau_\mathrm{gate}/\hbar)\right]
$
to the time-evolution operator is trivial.
Evaluating the effect of the operator $\mathbf{S}^2$ in the eigenbasis
(see also equation~\eqref{eq:H1_d} below), this is the case for gate times given by
\begin{equation}
  \tau_\mathrm{gate} = \frac{4\pi}{3} \frac{\hbar}{|J_\C|} \, m\,,
  \label{eq:taun}
\end{equation}
with the arbitrary integer $m>0$. It is important to note that
relation~\eqref{eq:taun} is independent of the spin direction of the
additional electron.  Equating conditions~\eqref{eq:taun} and
\eqref{eq:tswap}, we obtain the following relation between $J_1$ and
$J_\C$:
\begin{equation}
  \label{eq:gatingcondition}
  \frac{J_1}{|J_\C|} =  \mathrm{sgn} J_\C +
  \frac{3}{8} \frac{1-4n}{m}\,.
\end{equation}

\section{Realistic quantum-gate behavior}

\begin{figure}[ht]
  \centering
  \includegraphics[width=0.9\linewidth]{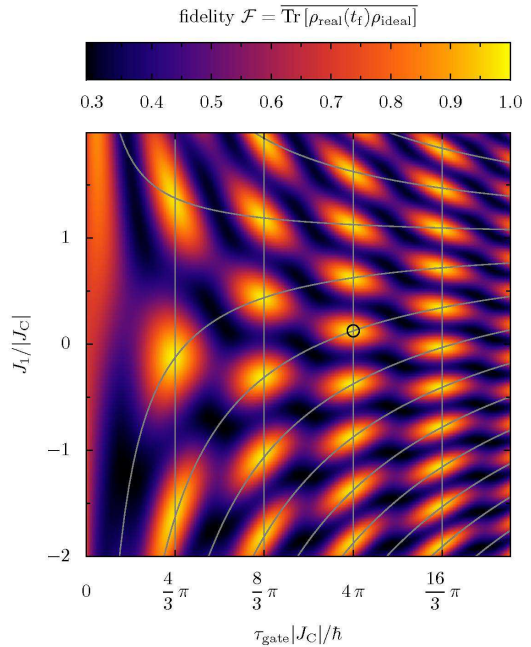}
  \caption{Average gate fidelity~$\mathcal{F}$ as a function of the
    gate time $\tau_\mathrm{gate}$ and the ratio $J_1/|J_\C|$ for a
    ferromagnetic $J_\C>0$. The conditions~\eqref{eq:tswap} and
    \eqref{eq:taun} are indicated by solid lines. The parameters are:
    $\hbar\Gamma = 5 |J_\C|$, $k_\mathrm{B}T = 0.001|J_\C|$, and
    $V_\mathrm{g,on}=-V_\mathrm{g,off}=15|J_\C|$ (measured from the
    charge-degeneracy point).}
  \label{fig:fidelity}
\end{figure}

In the preceding discussion we have assumed that the tunnelling
process of the electron is highly controlled, i.e., that we are
able to instantaneously switch between the $n_\C=0$ and $n_\C=1$
states. In reality, tunnelling is a probabilistic event occurring on a
mean time scale of the order of the inverse tunnelling rate. In order
to investigate to what extent this stochasticity affects our
analytical findings, we compare them with numerical results obtained
from a simulation of the incoherent quantum dynamics using the average
gate fidelity~$\mathcal{F}$ as figure of merit for the gate process
(see App.~\ref{sec:br} ).  In Fig.~\ref{fig:fidelity}, we show this
quantity as a function of the gate time and the ratio $J_1/|J_\C|$ for
the case of a ferromagnetic coupling $J_\C>0$ (the antiferromagnetic
case behaves very similarly).
We focus on the strong-coupling regime $\hbar\Gamma\gg |J_\C|$, where
tunnelling proceeds rapidly compared to the internal coherent
dynamics. As expected from our analytical considerations, the average
gate fidelity assumes maxima whenever both conditions~\eqref{eq:tswap}
and \eqref{eq:taun} are fulfilled.  We found the gate error
$1-\mathcal{F}$ to decrease with increasing tunnelling rate $\Gamma$. For
$\Gamma = 6 |J_\C|/\hbar$, for instance, we obtain a fidelity
$\mathcal{F}=0.99$ at the maximum indicated by the circle in
Fig.~\ref{fig:fidelity}.

\section{Readout process}

We now show that in a molecular system described by the
Hamiltonian~\eqref{eq:H}, it is possible to readout the quantum number
$S_0$ of the two outer spins by measuring the sequential tunnelling
current through the central dot. The current can be calculated using
the Bloch-Redfield equation approach (see App.~\ref{sec:br}).  For a
qualitative understanding it is sufficient to consider the allowed
transitions, i.e., non-zero matrix
elements~$\langle\alpha'|d^\dagger_{\C,\sigma}|\alpha\rangle$, between
the eigenstates of the two different charge sectors $n_\C=0$ and
$n_\C=1$.  The former are the eigenstates $|S_0, S_{0,z}\rangle$ of
the total spin
$\mathbf{S}_0=\mathbf{S}_\mathrm{L}+\mathbf{S}_\mathrm{R}$ of the two
outer spins.  The Hamiltonian~\eqref{eq:H1_t} of the molecule with an
additional electron can be readily diagonalized by rewriting it up to
an irrelevant constant as
\begin{equation}
  \label{eq:H1_d}
  H_1 = -\frac{J_\C}{2} \,\mathbf{S}^2 
  - \frac{J_1-J_\C}{2}\,
  \mathbf{S}_0^2
  \,.
\end{equation}
Thus, in the basis of the simultaneous eigenstates $|S, S_0,
S_z\rangle$ of $\mathbf{S}^2$, $\mathbf{S}_0^2$ and $S_z$, the
Hamiltonian in the $n_\C=1$ subsector is already diagonal. For the
matrix elements~$\langle
S',S'_0,S'_z|d^\dagger_{\C,\sigma}|S_0,S_{0,z}\rangle$, we then obtain the
selection rules $S'=S_0\pm 1/2$, $S'_0 = S_0$, $S'_z = S_{0,z}+\sigma$. In
particular, we note that the quantum number $S_0$ stays invariant
under a complete sequential-tunnelling cycle.  Since furthermore the
energy differences determining the tunnelling rates and thus the
current depend on the quantum number $S_0$, we find that the
(quasi-)stationary value of the current can be strongly dependent on
the initial preparation of $S_0$.

\begin{figure}[ht]
  \centering
  \includegraphics[width=0.9\linewidth]{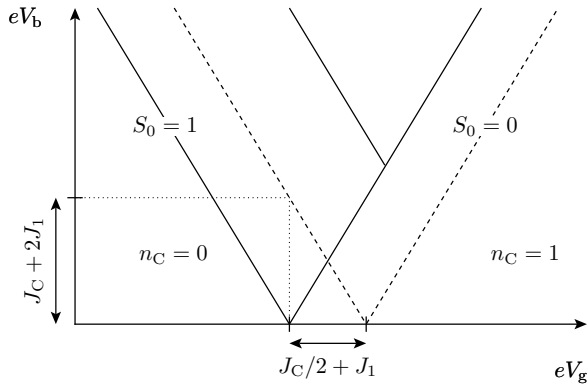}
  \caption{Sketch of ground and excited state tunnel
    transitions for $S_0=0$ (dashed line) and $S_0=1$ (solid line).}
  \label{fig:readout}
\end{figure}

This situation is depicted in Fig.~\ref{fig:readout}, where the
conductance peaks as a function of gate and bias voltage are shown
schematically for two different initial values of $S_0$.
From the figure we see that by starting initially at zero
bias~$V_\mathrm{b}=0$ in the $n_\C=0$ state, we first reach, upon
increasing~$V_\mathrm{b}$, the allowed ground-state transition and
thus the onset of the sequential tunnelling current for the triplet
$S_0=1$. On the other hand, for a singlet preparation~$S_0=0$,
sequential tunnelling will only be possible at a voltage which is
higher by $(J_\C+2 J_1)/e$. At low enough tunnelling rate and
temperature, i.e., $\hbar\Gamma, k_\mathrm{B}T \ll J_\mathrm{C} + 2
J_1$, it is possible to distinguish these two cases and hence to
measure the quantum number $S_0$. As discussed before,
for an independent control of the two parameters $V_\mathrm{g}$ and
$V_\mathrm{b}$ in Fig.~\ref{fig:readout}, a three-terminal geometry is
required. In the STM setup discussed here, one will only move along a line in
Fig.~\ref{fig:readout}, which, however, is sufficient for the readout
scheme, if the molecule is in the $n_\mathrm{C}=0$ charge sector at
$V_\mathrm{b}=0$.

Note that for an initial superposition with singlet and triplet
contributions, the readout process just described represents a
projective measurement. The first electron attempting to tunnel
determines the final outcome, i.e., the long-time current. The
average over many repeated measurements is described by the solution
of the master equation~\eqref{eq:master}.

We remark that the singlet-triplet readout process requires a
substantially smaller tunnel coupling than the one for the
exchange-controlled quantum-gate operation. There are several ways how
to achieve this: Obviously, one could increase the molecule-tip
distance which leads to an exponentially strong change in the
tunnelling rate. Still, doing so on a very short time-scale might be
technically challenging. Another possibility would be to use that the
tunnel barrier---and, thus, in turn the coupling strength---depends on
both the gate- and the bias-voltage.~\cite{Wolf1989a} Depending on the
details of the contact, this might be enough to achieve the required
change in the tunnel coupling.  Alternatively, one could transfer the
readout scheme to the so-called cotunnelling
regime,~\cite{Hirjibehedin2006a,Lehmann2006a} where transport proceeds
via virtual tunnelling on and off the molecule. Then, the onset of the
inelastic cotunnelling current is solely temperature-smeared, even in
the presence of a strong molecule-lead coupling.  For measuring $S_0$
one would then need to apply a magnetic field, which will lead to a
splitting of the triplet state $S_0=1$ only.  Thus, one will find only
for this state an inelastic cotunnelling step in the
conductance-voltage characteristics at the bias voltage corresponding
to the Zeeman energy.

\section{Ab-initio modelling and implementation requirements}

While the preceding discussion applies to any system described by a
Hamiltonian of the form~\eqref{eq:H}, we now return to the specific
case of the molecule [PMo$_{12}$O$_{40}$(VO)$_2$]$^{(q-)}$.  A high
redox flexibility is a characteristical feature of polyoxometalates,
where cyclic voltammetry experiments show one- or two-electron
reversible redox peaks, depending on the system and on the conditions.
In particular, four compounds based on the XMo$_{12}$O$_{40}$ (X = Si,
Ge, P) Keggin structure (either alone or vanadyl-bicapped) have been
recently found to show reversible two-electron electrochemical
processes.~\cite{Shi2006}
We thus extract the exchange coupling strengths for
different charge states~$N$ of the molecule (see App.~\ref{sec:par}).
Identifying even and odd values of $N$ with $n_\C=0$ and $n_\C=1$,
respectively, we obtain the parameters in the Hamiltonian~\eqref{eq:H}
as given in Table~\ref{par} for $0\le N\le6$; for higher electronic
populations $J_0$, $J_1$ and $J_C$ are of comparable size.  {
  \squeezetable
\begin{table}[h]
\begin{ruledtabular}
\begin{tabular}{cc}

\begin{tabular}{c|d|c}
$N$   & \multicolumn{1}{c|}{$J_0[\mathrm{meV}]$} & $E_\mathrm{gap}[\mathrm{meV}]$  \\
\hline
  0   &    0    & $>1000$  \\
  2   &   0.01  & $>100$   \\
  4   &   0.01  & $>100$   \\
  6   &   1.0   & $>100$   \\
\end{tabular}
\hspace{0.0cm}
&
\begin{tabular}{c|d|d|d|c}
$N$ &     \multicolumn{1}{c|}{$J_\C[\mathrm{meV}]$}     &
\multicolumn{1}{c|}{$\displaystyle\frac{J_1}{J_\C}$}  & 
\multicolumn{1}{c|}{$\alpha$}
& $\displaystyle\frac{E_\mathrm{gap}}{J_\C}$  \\
\hline
  1   &   1     &  0.01    &   <0.1     &     $>100$    \\
  3   &   1     &  0.1     &   <0.1     & $\simeq 1$    \\
  5   &   1     &  0.1     &   <0.1     &     $>10 $    \\
\end{tabular}
\end{tabular}
\end{ruledtabular}
\caption{Orders of magnitude of the exchange coupling strengths in the
  Hamiltonian~\eqref{eq:H} and estimates for the size of the
  correction terms, the asymmetry parameter
  $\alpha=(J_\mathrm{CL}-J_\mathrm{CR})/(J_\mathrm{CL}+J_\mathrm{CR})$
  and the gap, $E_\mathrm{gap}$, between the low-energy states described by the
  Hamiltonian~\eqref{eq:H} and further excited states, for different electronic
  populations $N$. The charging energy $U$ is of the order of $1\,\mathrm{eV}$.} \label{par}
\end{table}
}

It is worthwhile to consider how robust our predictions are against
deviations from the ideal experimental setup.  Specifically, (I) the
orientation of the molecules may be difficult to control, (II) its
electronic structure, and, specifically, its ideal symmetry, may be
altered upon deposition on a surface, (III) the molecule might be
affected by counter ions in the surrounding, and (IV) higher-energy
states not described by the Hamiltonian~\eqref{eq:H} might be
energetically accessible.

With respect to (I) we note that regardless of the orientation of the
molecule, the change in charge state is firmly expected to affect the
central Mo$_{12}$O$_{40}$ core and not the vanadyl groups, so we think
orientation will only have an additional effect on (II). As an
insulating thin layer separates the molecule from the metallic
substrate, the alteration in the electronic structure of the
polyoxometalate is expected to be minimal (details are discussed in
the App.~\ref{sec:par}).  The main effect on the low-energy physics will
be to introduce an asymmetry in the couplings between the left and
right vanadyl groups and the Keggin structure, $J_\mathrm{CL}$ and
$J_\mathrm{CR}$, respectively.  In Table~I, we give bounds for this
asymmetry, which can be quantified by the
parameter~$\alpha=(J_\mathrm{CL}-J_\mathrm{CR})/(J_\mathrm{CL}+J_\mathrm{CR})$.
Furthermore, we have verified that an asymmetry of up to $10\%$ does
still allow one to achieve a fidelity of $99\%$ in the regime $J_1\ll
J_\mathrm{CL}, J_\mathrm{CR}$ (for details see App.~\ref{sec:quant}). Concerning (III), cations needed to compensate the
negative charge of the polyoxometalate polyanions, as well as solvent
molecules, may be still present when the cluster is deposited.  This
feature may be a potentially important addition on (II) which could be
minimized using an electrospray system to choose the mass/charge ratio
corresponding to an isolated polyoxoanion.  Such techniques have
succesfully been applied to polyoxometalates and the peaks of the
isolated anions have been detected.\cite{Bonchio2003a,Mayer2004a}
Finally (IV), the full diagonalization of the effective Hamiltonian
also results in a set of excited states with energies $E\geq
E_\mathrm{gap}$ above those states described by the
Hamiltonian~\eqref{eq:H}: we also include these data in Table~I.

Having these points in mind, we see that different charge states are
usable.  If we consider for instance $N=4$ and $5$, one would need
molecule-lead coupling rates $\hbar\Gamma$ in the range of around
$0.1$ and $5 \, \mathrm{meV}$, for the gate and readout process,
respectively. For the gate procedure, voltage pulses of a precision on
the order of a few $\mathrm{ps}$ are required, which is not far from
what can be achieved with present technology.  In fact, if $N=0$ and
$N=1$ are found to be chemically stable in the experimental
environment, their parameters will be even more suitable for our
purposes by allowing $10$ times longer pulse times and yielding
$J_0=0$ as well as $E_\mathrm{gap}\gg J_0, J_\C$.

\section{Conclusions}

The present scheme defines a class of molecular systems: those where two
localized spins are coupled to a redox-active unit.  The physically relevant
figures of merit are included in Table~\ref{par}; the chemical desiderata are
stability, facility of deposition on surfaces, and possibility of controlled
oligomerization.  Indeed, scalability of the present scheme requires covalent
bonding and directed self-assembly of these logical building blocks.

Molecules of this class can be found in several chemical families,
besides bicapped polyoxometalates.  Tetrathiofulvalene derivatives,
polypyrrol, porphyrines/phthalocyanines, fullerene and single-wall
nanotubes are a few among a plethora of organic systems which can
reversibly lose or gain one electron, can be substituted with radical
groups, and have a rich and well-controlled chemistry.  For example,
theoretical calculations on phthalocyanines point to the possibility of
the chemical tailoring of these molecules to enable exchange couplings
in the range required for the gate scheme described
here~\cite{Shultz1999}.

While for an initial experimental realization with a single molecule,
an STM contacting scheme as discussed in the present article enables
the best control and is clearly favorable, a scalable method for a
molecular monolayer could be based, e.g., on a crossbar architecture
which already with current technology reaches very high
densities.\cite{Green2007a}

In conclusion, we have proposed an experimental setup for
single-molecule all-electric two-qubit gate and readout which is
within reach of current technology.  We have exemplified our scheme
using a mixed-valence polyoxometalate, for which the model
parameters have been calculated using an \textit{ab initio} approach.
The general principles
behind our proposal also apply to different classes of molecular
systems. The chemical design, synthesis and characterization of such
systems should open new routes to molecular spin-qubit quantum
computing.

\begin{acknowledgments}

We thank B. Coish for discussions. Financial support by the EU RTN
QuEMolNa, the EU NoE MAGMANet, the NCCR Nanoscience, the Swiss NSF, the
Spanish MEC (MAT2004-3849) and the Generalitat Valenciana is
acknowledged.

\end{acknowledgments}

\appendix

\section{Bloch-Redfield-equation approach and average gate fidelity}

\label{sec:br}

For the description of the molecular dynamics in the presence of the
molecule-lead coupling we employ a Bloch-Redfield-equation formalism.
This approach, which is valid in the sequential-tunnelling regime,
i.e., to lowest order in the tunnelling rates $\Gamma_\ell$, includes the
full coherent quantum dynamics of the spin
states~\cite{Engel2001a,Kohler2005a} as described by the density
matrix elements $\varrho_{\alpha\beta} = \mathrm{Tr}_\leads \langle
\alpha|\varrho|\beta\rangle$ after tracing out the leads.  In the
usual Born-Markov approximation, we obtain
\begin{equation}
  \label{eq:master}
  \begin{split}
  \dot{\varrho}_{\alpha\beta} = &
  -\I \omega_{\alpha\beta} \varrho_{\alpha\beta}
  + \frac{1}{2}
  \sum_{\ell \alpha' \beta'} \!
  \bigg\{\!\!
  \left[W^\ell_{\beta' \beta \alpha \alpha'}  + (W^\ell_{\alpha' \alpha \beta \beta'})^\ast\right]
  \varrho_{\alpha'\beta'}\\
  &\quad-
  W^\ell_{\alpha \beta'\beta'\alpha'}\,
  \varrho_{\alpha'\beta}
  -
  (W^\ell_{\beta \alpha'\alpha'\beta'})^\ast\,
  \varrho_{\alpha\beta'}
  \bigg\}\,,
  \end{split}
\end{equation}
where we have introduced the frequencies
$\omega_{\alpha\beta}=(E_\alpha-E_\beta)/\hbar$ and the transition
rates $W^\ell_{\beta' \beta \alpha \alpha'} = W^{\ell+}_{\beta' \beta
  \alpha \alpha'} + W^{\ell-}_{\beta' \beta \alpha \alpha'}$ due to
tunnelling across a molecule-lead contact $\ell$.  The rates for
tunnelling of an electron on and off the molecule assume the form
$W^{\ell+}_{\beta' \beta \alpha \alpha'} = \Gamma_{\ell} \,
f(E_\alpha-E_{\alpha'}-\mu_\ell) \sum_{\sigma} \langle
\beta'|d_{\C,\sigma}|\beta\rangle\!\langle
\alpha|d_{\C,\sigma}^\dagger|\alpha'\rangle$ and $W^{\ell-}_{\beta'
  \beta \alpha \alpha'} = \Gamma_{\ell} \,
[1-f(E_{\alpha'}-E_{\alpha}-\mu_\ell)] \sum_{\sigma} \langle
\beta'|d^\dagger_{\C\sigma}|\beta\rangle\!\langle
\alpha|d_{\C,\sigma}|\alpha'\rangle$, respectively.  Here, we assume
that the lead electrons are described by a Fermi distribution
$f(\epsilon-\mu_\ell) = \{1+\exp[(\epsilon-\mu_\ell)/kT]\}^{-1}$ at
temperature~$T$ and electro-chemical potential~$\mu_\ell$.  The
tunnelling current across contact $\ell$ can be determined from the
density matrix elements via $ I_\ell = e\, \mathrm{Re} \sum_{\alpha
  \alpha' \beta} \big( W^{\ell-}_{\beta \alpha' \alpha' \alpha} -
W^{\ell+}_{\ell \beta \alpha' \alpha' \alpha}
\big)\varrho_{\alpha\beta}$.

Using the Bloch-Redfield equation approach, we are able to simulate a
realistic gate cycle. This allows us to calculate the average gate
fidelity~\cite{Poyatos1997a} $\mathcal{F}=\overline{\mathrm{Tr}
  \left[\rho_\mathrm{real}(t_\mathrm{f}) \rho_\mathrm{ideal}\right]}$
as quantitative measure for the gate quality.  Here, the overline
indicates the average over 16 unentangled input states
$|\Psi_i\rangle_\mathrm{L} |\Psi_j\rangle_\mathrm{R}$, where
$|\Psi_1\rangle=|{\uparrow}\rangle$,
$|\Psi_2\rangle=|{\downarrow}\rangle$,
$|\Psi_3\rangle=(|{\uparrow}\rangle + |{\downarrow}\rangle)/\sqrt{2}$,
and $|\Psi_4\rangle=(|{\uparrow}\rangle + \I
|{\downarrow}\rangle)/\sqrt{2}$, $\rho_\mathrm{ideal}$ is the ideal
result of the $\sqrt{\mathrm{SWAP}}$ operation as defined above and
$\rho_\mathrm{real}(t_f)$ is the state of the system according to the
quantum dynamics~\eqref{eq:master} at the final time~$t_f$.

\section{Evaluation of exchange parameters}

\label{sec:par}

For the evaluation of the exchange coupling constants and to check the
validity of our numerical assumptions, we used an effective
Hamiltonian approach. We considered the main one- and two-center
phenomena in mixed-valence systems, namely magnetic exchange, electron
transfer, Coulomb repulsion and orbital energy, which were
parametrized for [PMo$_{12}$O$_{40}$(VO)$_2$]$^{q-}$ (details will be
published elsewhere).  Thus, we diagonalized a 14-site
$t$--$J$--$V$--$\epsilon$ effective Hamiltonian for different
electronic populations. We projected this full energy level scheme
into a subsystem containing only the localized spins in the
[V$^{IV}$O] groups, plus, for odd values of $N$, an unpaired spin in
the Mo$_{12}$O$_{40}$ Keggin structure.

For $N$ \textit{even}, $J_0$ is given by the energy difference between the
lowest singlet and triplet states, while $E_\mathrm{gap}$ is the gap
to the next energy level.  For $N$ \textit{odd}, the evaluation of
$E_\mathrm{gap}$ is equally direct. Obtaining the exchange parameters
is also straightforward when a symmetric molecular structure is assumed:
$J_\C$ and $J_1$ can then be readily calculated from the energy
differences between the first doublet and quadruplet states.
When deviations from symmetry are considered, the
determination of the three exchange parameters $J_\C$,
$J_\mathrm{CL}$ and $J_\mathrm{CR}$ requires the analysis of the
wavefunctions.

The spatially-dependent microscopic parameters $t$, $J$, $V$, and
$\epsilon$ were obtained from ab-initio calculations of molecular
subclusters, which accounts for inhomogeneities in the molecule. The
experimental setup is expected to imply small variations in bond
distances and angles, as well as a different charge distribution in
the surrounding of the molecule. In order to assess how these
perturbations affect the parameters in the effective
Hamiltonian~\eqref{eq:H}, the calculations were repeated for a
reasonable range of deviations ($\pm 50\%$ in $J$ and $t$, and $\pm
1$eV in $\epsilon$). Except for the case with $N=3$, where the excited
states not contained in the Hamiltonian~\eqref{eq:H} lie close in
energy, the parameters of our interest proved to be robust enough
within the orders of magnitude indicated in Table~\ref{par}.

\section{Characteristic gate quantifiers}

\label{sec:quant}

Besides the average gate fidelity discussed in the main text, one can
consider further quantities for assessing the quality of a
quantum-gate operation~\cite{Poyatos1997a}. One is the so-called gate
purity
\begin{equation}
  \label{eq:s1}
  \mathcal{P} = \overline{\mathrm{Tr} \left[\rho_\mathrm{real}(t_\mathrm{f})^2\right]}\,,
\end{equation}
which is a measure for the decoherence of the gate. Here, the average
is over all 16 unentangled input states $|\psi_\mathrm{ue}(t_0)\rangle
= |\Psi_i\rangle_\mathrm{L}|\Psi_j\rangle_\mathrm{R}$, $i,j=1,\dots4$
(see App.~\ref{sec:br} for the definition of the $|\Psi_i\rangle$). For
an ideal $\sqrt{\mathrm{SWAP}}$ operation it reaches unity.  Another
measure is the entanglement capability~$\mathcal{C}$ which is defined
as the smallest eigenvalue of the partial transpose of the density
matrix~\cite{Peres1996a} $\rho_\mathrm{real}(t_\mathrm{f})$ obtained
for all 16 unentangled input states just defined. Ideally, it assumes
a value of $-1/2$.  Finally, the quantum degree of the gate is given
by
\begin{equation}
  \label{eq:s3}
  \mathcal{Q} = \, \max_{|\psi_\mathrm{me}\rangle, \rho_\mathrm{ue}(t_0)}
                  \langle\psi_\mathrm{me}| \rho_\mathrm{real}(t_\mathrm{f}) |\psi_\mathrm{me}\rangle\,,
\end{equation}
i.e., as the maximum of the overlap between all possible output states
obtained for all 16 unentangled input states and all maximally
entangled states~$|\psi_\mathrm{me}\rangle$, viz, the four Bell
states $(|{\uparrow}\rangle_\mathrm{L}|{\uparrow}\rangle_\mathrm{R}
\pm
|{\uparrow}\rangle_\mathrm{L}|{\uparrow}\rangle_\mathrm{R})/\sqrt{2}$
and $(|{\downarrow}\rangle_\mathrm{L}|{\uparrow}\rangle_\mathrm{R} \pm
|{\uparrow}\rangle_\mathrm{L}|{\downarrow}\rangle_\mathrm{R})/\sqrt{2}$.
In the optimal case, the quantum degree reaches $1/2$ for the
$\sqrt{\mathrm{SWAP}}$ gate.

\begin{figure*}[ht]
  \includegraphics[width=0.9\textwidth]{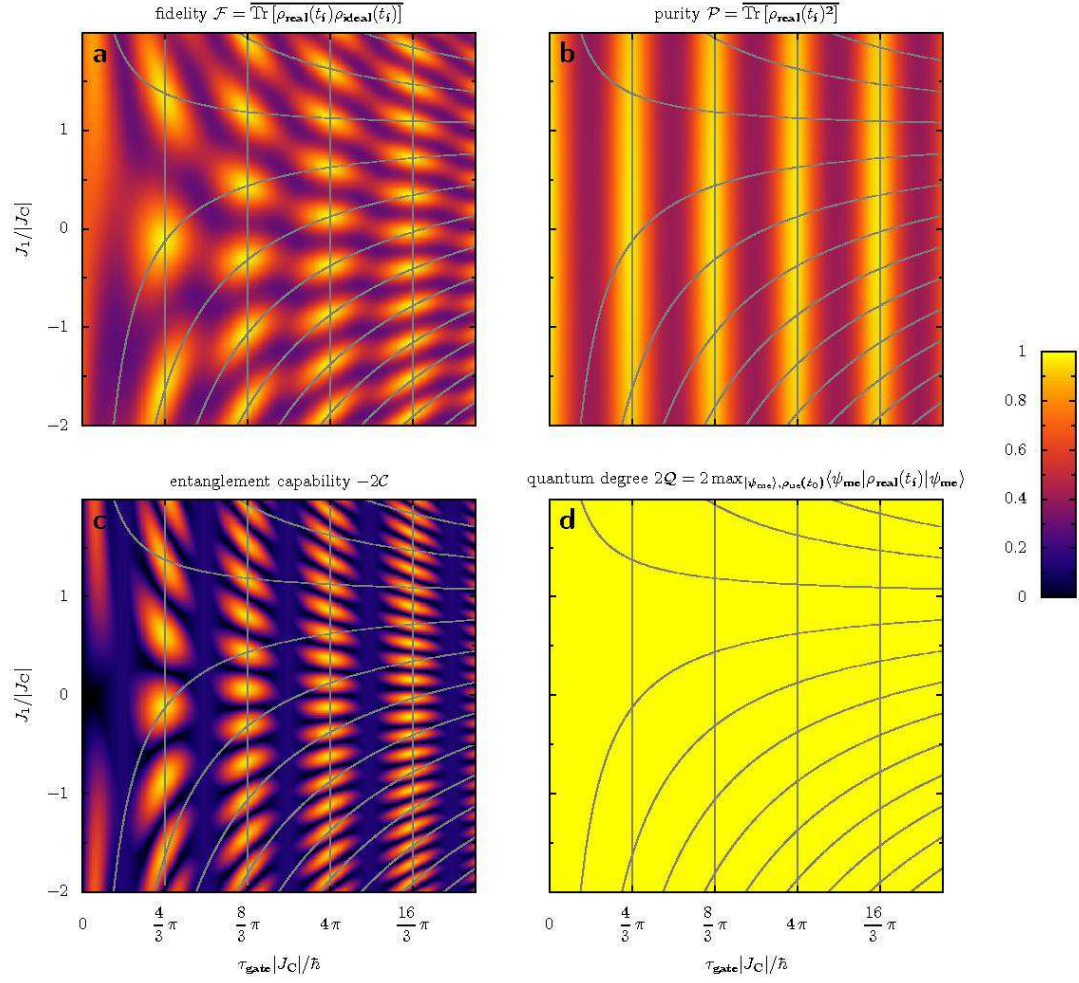}
  \caption{Four average gate quantifiers (a) gate fidelity (b) purity
    (c) entanglement capability, and (d) quantum degree each
    normalized to the maximal value achievable for an ideal
    $\sqrt{\mathrm{SWAP}}$ operation as a function of the gating
    time~$\tau_\mathrm{gate}$ and the ratio $J_1/|J_\mathrm{C}|$. The
    parameters are as in Fig.~3.  The conditions (3) and (4) are
    indicated by solid lines. Note that the color scale for the gate
    fidelity is different from the one  used in Fig.~3.}
  \label{fig:S1}
\end{figure*}

In Fig.~\ref{fig:S1}, we show these four gate quantifiers as a
function of the gating time~$\tau_\mathrm{gate}$ and the ratio
$J_1/|J_\mathrm{C}|$. The gate fidelity~$\mathcal{F}$ has already been
discussed in the main text. We observe that the gate
purity~$\mathcal{P}$ assumes almost unity for the stroboscopic gating
times~(4). This means that at these times, the additional
spin~$\mathbf{s}_\mathrm{C}$ and the spin $\mathbf{S}_0 =
\mathbf{S}_\mathrm{L}+\mathbf{S}_\mathrm{R}$ become disentangled again,
such that one recovers a pure state when removing the additional
electron. Furthermore, we find that all gate quantifiers except for
the gate fidelity overestimate the quality of the gate: They almost
assume their perfect values even for parameters where the gate
fidelity is much below unity. The quantum degree~$\mathcal{Q}$ turns
out to be completely independent of the gating time and the
exchange-coupling ratio.

\begin{figure*}[ht]
  \includegraphics[width=0.9\textwidth]{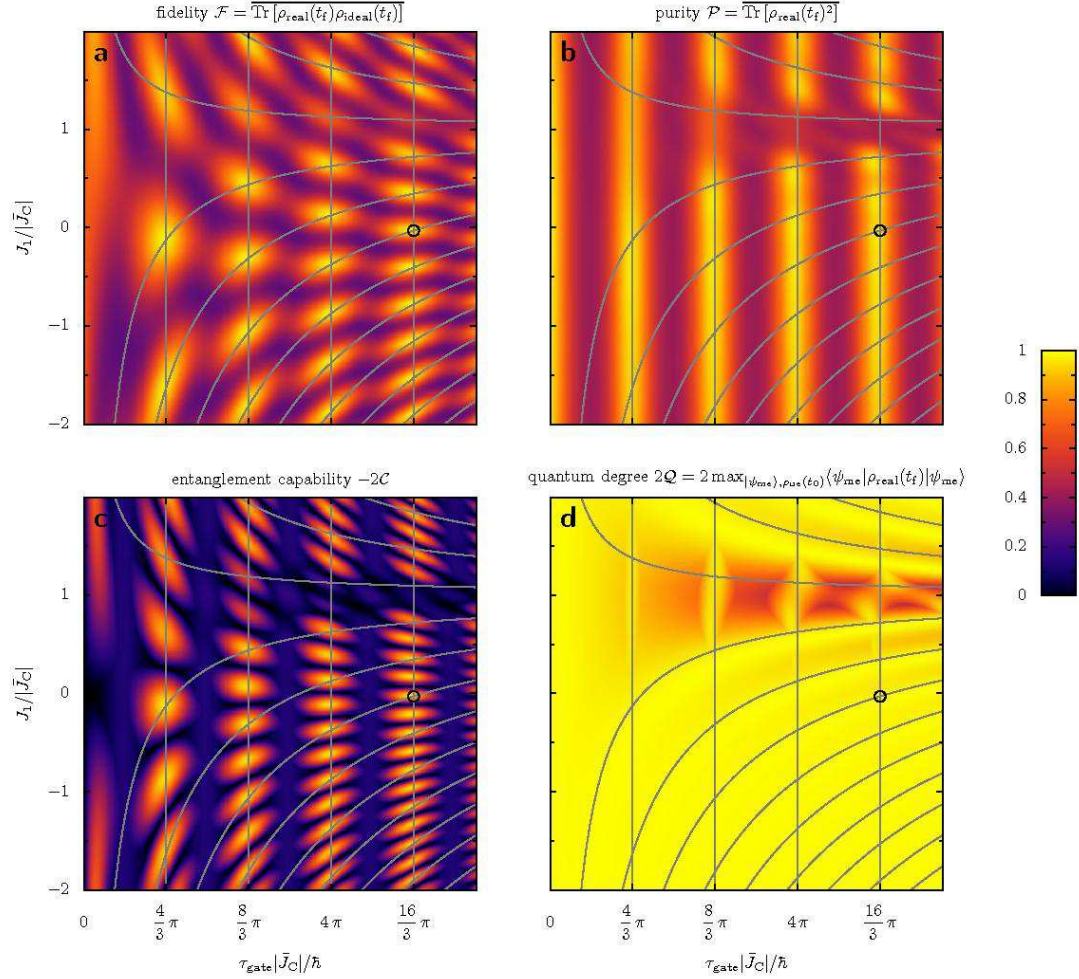}
  \caption{Four average gate quantifiers (a) gate fidelity (b) purity
    (c) entanglement capability, and (d) quantum degree each
    normalized to the maximal value achievable for an ideal
    $\sqrt{\mathrm{SWAP}}$ operation as a function of the gating
    time~$\tau_\mathrm{gate}$ and the ratio $J_1/|\bar J_\mathrm{C}|$,
    where $\bar J_\mathrm{C} = (J_\mathrm{CL}+J_\mathrm{CR})/2$ is the
    average exchange coupling between the Keggin core and the
    left and right vanadyl groups. An asymmetry
    $\alpha=(J_\mathrm{CL}-J_\mathrm{CR})/(J_\mathrm{CL}+J_\mathrm{CR})=0.1$
    of these exchange couplings is assumed. The other parameters are
    as in Fig.~3.  The conditions (3) and (4) are indicated by solid
    lines.}
  \label{fig:S2}
\end{figure*}

The situation in the presence of an
asymmetry~$\alpha=(J_\mathrm{CL}-J_\mathrm{CR})/(J_\mathrm{CL}+J_\mathrm{CR})$
between the couplings $J_\mathrm{CL}$ and $J_\mathrm{CR}$ of the
central Keggin core to the left and right, respectively, vanadyl
groups is shown in Fig.~\ref{fig:S2}. We have chosen the upper bound
$\alpha=0.1$ from the results of our \textit{ab initio} calculations
(cf.\ Table I). We find for the gating times shown in
Fig.~\ref{fig:S2} that this noticeable aberration from perfect
symmetry does not strongly affect the gate quality for small and/or
negative values of the ratio $J_1/|\bar J_\mathrm{C}|$, where $\bar
J_\mathrm{C} = (J_\mathrm{CL}+J_\mathrm{CR})/2$ is the average
coupling of the central spin to the left and right spins.
In particular, we still find a fidelity better than $\mathcal{F}=0.99$ at the
maximum indicated by the circles in Fig.~\ref{fig:S2} for a tunnel
rate $\Gamma=11 |\bar J_\mathrm{C}|/\hbar$.

\end{document}